\documentclass[12pt]{article}

\usepackage{amsmath}
\usepackage{amssymb}
\usepackage{slashed}
\usepackage[numbers,sort&compress]{natbib}
\usepackage{hyperref}
\usepackage[normalem]{ulem}
\usepackage{cleveref}
\crefname{equation}{Eq.}{Eqs.}
\crefname{figure}{Fig.}{Figs.}
\crefname{table}{Table}{Tables}
\crefname{section}{Section}{Sections}

\newcommand\scalemath[2]{\scalebox{#1} {\mbox{\ensuremath{\displaystyle #2}}}}

\usepackage[letterpaper,margin=1in,bottom=1in]{geometry}
\usepackage{float} % allows forcing plot and table locations with [H]
\usepackage{parskip} % skip spaces between paragraphs
\usepackage{tabulary} % table environment with auto word wrapping
\usepackage{xcolor} % color for tracking edits
\usepackage{soul} % strikeout for tracking edits
\usepackage{subfigure}
\usepackage{graphicx}
\usepackage[section]{placeins} % keep figures inside their sections

\def\lhc2{LHC~Run~II}

\usepackage{cleveref}

\newcommand{\code}[1]{\texttt{#1}}

\def\.4{\vspace{-.5cm}}
\newcommand{\ifb}{~\textrm{fb}^{-1}}

\def\beq{\begin{equation}}
\def\be{\begin{equation}}
\def\beqn{\begin{eqnarray}}
\def\ee{\end{equation}}
\def\eeq{\end{equation}}
\def\eeqn{\end{eqnarray}}

\def\st{Stueckelberg~}

\author{
Amin Aboubrahim\footnote{Email: a.abouibrahim@northeastern.edu}~\ and 
Pran Nath\footnote{Email: p.nath@northeastern.edu}\\~\\
\textit{Department of Physics, Northeastern University,
Boston, MA 02115-5000, USA}
}

\title{LHC phenomenology with hidden sector dark matter:\\  a long-lived stau and heavy Higgs in an observable range}

\date{}

\begin{document}
\maketitle

\textbf{Abstract: } 
The presence of a hidden sector with very weak interactions with the standard model has significant implications on LHC signatures.
In this work we discuss LHC phenomenology with the inclusion of a hidden sector by a $U(1)$ extension of MSSM/SUGRA.
We consider both kinetic mixing and \st mass mixing between the $U(1)$ gauge field of the hidden sector and $U(1)_Y$ of the visible sector.
Such a model has an extended parameter space. We consider here two limited regions of this parameter space. 
In the first case we consider a $U(1)$ gauge field along with chiral fields needed for the \st mechanism to operate and discuss the mixing between the 
hidden and the visible sectors.  Here if the stau is the lightest sparticle in the MSSM sector  and the neutralino of the hidden sector is the LSP of the full system and a dark matter candidate,
the stau can be long-lived and decay inside an LHC detector tracker. 
In the second case we include extra vectorlike matter in the hidden sector which can give rise to a Dirac fermion in addition to the two neutralinos in 
the hidden sector. The neutralino sector now has six neutralinos and we assume that the lightest of these is the LSP and is higgsino-like.
In this case the dark matter is constituted of a Majorana and a Dirac fermion, and a small $\mu$ leads to heavy Higgs boson masses which reside in
the observable range of HL-LHC and HE-LHC.

\vspace{1cm}

\textit{Talks presented at the 2019 Meeting of the Division of Particles and Fields of the American Physical Society (DPF2019), July 29 - August 2, 2019, Northeastern University, Boston, C1907293} 
 
\newpage

\section{Introduction}\label{sec:intro}

Within supersymmetry (SUSY) and supergravity (SUGRA) unified models (for a review see, e.g.~\cite{Nath:2016qzm}), the mass of the standard model (SM) Higgs boson is expected to be below 130 GeV~\cite{Akula:2011aa,Arbey:2012dq,Akula:2012kk} which is exactly what experiments at the Large Hadron Collider (LHC) have measured, where a SM-like Higgs boson was observed at $\sim 125$ GeV~\cite{Aad:2012tfa,Chatrchyan:2012ufa} lending support to SUSY. As is well known the Higgs boson mass at $\sim 125$ GeV requires a large loop correction within the minimal supersymmetric standard model (MSSM) which in turn implies that the size of weak scale supersymmetry is large, lying in the several TeV region. This also explains why SUSY has not been  observed at accelerators thus far. In this work we consider a $U(1)$ extension of MSSM/SUGRA and the resulting phenomenology. This phenomenology leads to 
new phenomena observable at the LHC in different parts of its parameter space. We consider two specific parts, one where 
stau is the lightest supersymmetric particle (LSP) of the MSSM sector and the LSP  of the extended sector is a hidden sector neutralino which is a dark matter candidate.
In the second investigation we consider the case where a Majorana neutralino in the MSSM sector and a Dirac fermion in the 
hidden sector form two components of dark matter. We give now further details of the model.
 
 The extension proposed in this work brings in an additional vector superfield with particle content of 
$B_\mu', \lambda_{X}$ where $B_\mu'$ is the new gauge boson and $\lambda_{X}$ is its gaugino superpartner.
 The $U(1)_X$ can mix with hypercharge  $U(1)_Y$ via kinetic mixing~\cite{Holdom:1985ag, Holdom:1991}. 
 Additionally with Stueckelberg mass mixing of 
 $U(1)_X$ and $U(1)_Y$ one brings in a chiral superfield which contains a Weyl fermion $\psi$~\cite{Kors:Nath,st-mass-mixing,Feldman:2007wj}.
  After electroweak symmetry breaking the above leads to a $6\times 6$ neutralino mass matrix, where the additional two neutralinos reside in the hidden
 sector with highly suppressed couplings to the visible sector. If the next-to-lightest supersymmetric particle (NLSP) of the extended model is a stau and the LSP is the lightest neutralino of the hidden sector then a stau decaying to the hidden sector neutralino is long-lived and since it is charged will leave a track in the inner detectors (ID) of the ATLAS and CMS experiments.

In the second class of extended $U(1)$ models we consider we 
add hidden sector matter which are neutral under $U(1)_Y$ while charged only under $U(1)_X$. Such a setup is useful when dealing with a DM candidate in the visible sector with small relic abundance. Thus to saturate the relic density to the experimentally measured value, one needs at least one extra component which could be a Dirac fermion of the hidden sector matter. A higgsino-like LSP (obtained when the $\mu$ parameter is small) of the visible sector fits the description of an LSP with small relic density. Another  consequence of small $\mu$ is the presence of MSSM heavy Higgs bosons which can be under 1 TeV and accessible to the LHC. 

In the first part of this work we explore the possibility of observing charged long-lived particles (LLP) within the 
 framework of supergravity  grand unified model
 with an extended $U(1)_X$ sector including both the gauge kinetic mixing and the Stueckelberg mass mixing. In the second part we focus on SUGRA models on the hyperbolic branch with a small $\mu$ (of order the electroweak scale) where the LSP is higgsino-like. Indeed within radiative breaking of the electroweak symmetry higgsino-like dark matter can arise naturally on the hyperbolic branch when $\mu$ is small~\cite{Chan:1997bi,Chattopadhyay:2003xi,Akula:2011jx} (for related works, see, e.g.,~\cite{Feng:1999mn,Baer:2003wx,Feldman:2011ud,Ross:2017kjc}). 
Such models are highly constrained because the DM-proton scattering cross-section is large and mostly ruled out especially for WIMPs in the few hundreds of GeV in mass. However, such models can be viable if DM is multicomponent with the higgsino-like DM contributing only a fraction of the relic density with the rest made up from other sources. Here we discuss a two-component DM model where one component is the higgsino (a Majorana fermion) of the visible sector while the other component arises from the hidden sector and is a Dirac fermion~\cite{Feldman:2010wy}. One of the important consequences of the two-component model is the prediction of a relatively light CP odd Higgs (in the range of few hundred GeV) which can be within the reach of high luminosity LHC (HL-LHC) and high energy LHC (HE-LHC)~\cite{Benedikt:2018ofy,Zimmermann:2018koi}. In this work we carry out a detailed analysis of the integrated luminosities needed for the observation of this low-lying Higgs. Its observation would lend support to the higgsino nature of the LSP and the multi-component nature of dark matter. For additional work on the prospects of SUSY discovery at HL-LHC and HE-LHC, see Refs.~\cite{Aboubrahim:2018tpf,Aboubrahim:2018bil,Aboubrahim:2017wjl,Aboubrahim:2017aen,Han:2019grb,Baer:2018hpb}.   

\section{The model}\label{sec:model}

As discussed above  we consider an extension  of the standard model gauge group by an additional abelian gauge group $U(1)_X$ of gauge coupling strength $g_X$ where the MSSM particle spectrum of the visible sector are assumed neutral under $U(1)_X$. Thus the abelian gauge sector of the extended model contains two vector superfields, a vector superfield $B$ associated with the hypercharge gauge group $U(1)_Y$, a vector superfield $C$ associated with the hidden sector gauge group $U(1)_X$, and a chiral scalar superfield $S$. 

The gauge kinetic energy  sector of the model is given by
\begin{equation}
\mathcal{L}_{\rm gk}=-\frac{1}{4}(B_{\mu\nu}B^{\mu\nu}+C_{\mu\nu}C^{\mu\nu})-i\lambda_B\sigma^{\mu}\partial_{\mu}\bar{\lambda}_B-i\lambda_{C}\sigma^{\mu}\partial_{\mu}\bar{\lambda}_{C}+\frac{1}{2}(D^2_B+D^2_{C}).
\label{kinetic-1}
\end{equation} 
We invoke gauge kinetic mixing between the $U(1)_X$ and $U(1)_Y$ sectors via terms of the form
\begin{equation}
-\frac{\delta}{2}B^{\mu\nu}C_{\mu\nu}-i\delta(\lambda_{C}\sigma^{\mu}\partial_{\mu}\bar{\lambda}_B+\lambda_{B}\sigma^{\mu}\partial_{\mu}\bar{\lambda}_{C})+\delta D_B D_{C}.
\label{kinetic-2}
\end{equation}
As a result of Eq.~(\ref{kinetic-2}) the hidden sector interacts with the MSSM fields through the small kinetic mixing parameter $\delta$. The kinetic terms in Eq.~(\ref{kinetic-1}) and Eq.~(\ref{kinetic-2})  can be diagonalized by the transformation
\beqn
\left(\begin{matrix} B^{\mu} \cr 
C^{\mu} 
\end{matrix}\right) = \left(\begin{matrix} 1 & -s_{\delta} \cr 
0 & c_{\delta} 
\end{matrix}\right)\left(\begin{matrix} B'^{\mu} \cr 
C'^{\mu} 
\end{matrix}\right), 
\label{rotation}
\eeqn  
where $c_{\delta}=1/(1-\delta^2)^{1/2}$ and $s_{\delta}=\delta/(1-\delta^2)^{1/2}$. 

Alongside gauge kinetic mixing, we assume a Stueckelberg mass mixing between the $U(1)_X$ and $U(1)_Y$ sectors so that~\cite{Kors:Nath}
\begin{equation}
\mathcal{L}_{\rm St}=\int d\theta^2 d\bar{\theta}^2(M_1 C+M_2 B+S+\bar{S})^2.
\label{lag}
\end{equation}  

The matter sector of the model consists of the visible sector chiral superfields denoted by $\Phi_i$ where $i$ runs over all quarks, squarks, leptons, sleptons, Higgs and Higgsino fields of the MSSM which are neutral under $U(1)_X$ and hidden sector chiral superfields denoted by $\Psi_i$ and are neutral under $U(1)_Y$. The Lagrangian for the matter interacting with the $U(1)$ gauge fields is given by
\begin{equation}
\mathcal{L}_{\rm m}=\int d^2\theta d^2\bar{\theta}\sum_i \left[\bar{\Phi}_i e^{2g_Y Y B+2g_X X C}\Phi_i + \bar{\Psi}_i e^{2g_Y Y B+2g_X X C}\Psi_i\right],
\end{equation} 
where $Y$ is the $U(1)_Y$ hypercharge and $X$ is the $U(1)_X$ charge. The minimal particle content of the hidden sector consists of a left chiral multiplet $\Psi=(\phi, f, F)$ and a charge conjugate $\Psi^c=(\phi', f', F')$ so that $\Psi$ and $\Psi^c$ carry opposite $U(1)_X$ charge and hence constitute an anomaly-free pair. The Dirac field $\psi$ formed by $f$ and $f'$ has a mass $M_{\psi}$ arising from the term $M_{\psi}\Psi\Psi^c$ in the superpotential. Following SUSY breaking, the scalar fields of the hidden sector acquire soft masses equal to $m_0$ (the universal scalar mass of the visible sector) and the additional Dirac mass such that
\begin{equation}
m^2_{\phi}=m^2_0+M^2_{\psi}=m^2_{\phi'}.	
\label{mphi}
\end{equation} 
It is convenient from this point on to introduce Majorana spinors $\psi_S$, $\lambda_X$ and $\lambda_Y$ so that   
 \begin{equation}
  \psi_S =
  \begin{pmatrix}
    \chi_{\alpha}  \\
    \bar{\chi}^{\dot{\alpha}} 
  \end{pmatrix},\quad
  \lambda_X=
  \begin{pmatrix}
    \lambda_{C\alpha}  \\
    \bar{\lambda}^{\dot{\alpha}}_{C} 
  \end{pmatrix},\quad
  \lambda_Y=
  \begin{pmatrix}
    \lambda_{B\alpha}  \\
    \bar{\lambda}^{\dot{\alpha}}_{B}
  \end{pmatrix}.
  \label{spinors}
\end{equation}
In addition to the MSSM soft SUSY breaking terms, we add new soft terms related to the additional fields
\begin{equation}
\Delta\mathcal{L}_{\rm soft}=-\left(\frac{1}{2}m_X\bar{\lambda}_X\lambda_X+M_{XY}\bar{\lambda}_X\lambda_Y\right)-\frac{1}{2}m^2_{\rho}\rho^2,
\end{equation}
where $m_X$ is the $U(1)_X$ gaugino mass and $M_{XY}$ is the $U(1)_X-U(1)_Y$ mixing mass. \\

After electroweak symmetry breaking,  $\psi_S$ and $\lambda_X$ mix with the MSSM gauginos and higgsinos to form a $6 \times 6$ neutralino mass matrix. We choose as basis $(\psi_S,\lambda_X,\lambda_Y,\lambda_3,\tilde h_1, \tilde h_2)$  where the first two fields arise from the extended sector and the last four, i.e., 
 $\lambda_Y, \lambda_3, \tilde h_1, \tilde h_2$ are the gaugino and the
 higgsino fields of the MSSM sector. Using the transformation of Eq.~(\ref{rotation}) we rotate into the new basis $(\psi_S, \lambda'_X, \lambda'_Y,\lambda_3,\tilde h_1, \tilde h_2)$ so that the $6\times 6$ neutralino mass matrix takes the form
\beqn
\scalemath{0.9}{
\left(\begin{array}{cc|cccc} 0 & M_1 c_{\delta}-M_2 s_{\delta} & M_2 & 0 & 0 & 0 \cr
M_1 c_{\delta}-M_2 s_{\delta} & m_X c^2_{\delta}+m_1 s^2_{\delta}-2M_{XY}c_{\delta}s_{\delta} & -m_1 s_{\delta}+M_{XY}c_{\delta} & 0 & s_{\delta}c_{\beta}s_W M_Z & -s_{\delta}s_{\beta}s_W M_Z \cr
\hline
M_2 & -m_1 s_{\delta}+M_{XY}c_{\delta} & m_1 & 0 & -c_{\beta}s_W M_Z & s_{\beta}s_W M_Z \cr
0 & 0 & 0 & m_2 & c_{\beta}c_W M_Z & -s_{\beta}c_W M_Z \cr
0 & s_{\delta}c_{\beta}s_W M_Z & -c_{\beta}s_W M_Z & c_{\beta}c_W M_Z & 0 & -\mu \cr
0 & -s_{\delta}s_{\beta}s_W M_Z & s_{\beta}s_W M_Z & -s_{\beta}c_W M_Z & -\mu & 0 \cr
\end{array}\right)},
\eeqn  
where 
$s_{\beta}\equiv\sin\beta$, $c_{\beta}\equiv\cos\beta$, $s_W\equiv\sin\theta_W$, $c_W\equiv\cos\theta_W$ with $M_Z$ being the $Z$ boson mass. 
 We label the mass eigenstates as 
 \begin{equation}
 \tilde\xi^0_1, ~\tilde\xi^0_2; ~\tilde \chi_1^0, ~\tilde \chi_2^0, ~\tilde \chi_3^0, ~\tilde \chi_4^0\,,
 \end{equation}
where $\tilde\xi^0_1$ and $\tilde\xi^0_2$ belong to the hidden sector and mix with the usual MSSM neutralinos. In the limit of small mixings between the hidden and the MSSM sectors the masses of the hidden sector neutralinos are given by 
\begin{equation}
m_{\tilde\xi^0_1}=\sqrt{M_1^2+\frac{1}{4}\tilde m^2_X}-\frac{1}{2}\tilde m_X, \quad \text{and} \quad m_{\tilde\xi^0_2}=\sqrt{M_1^2+\frac{1}{4}\tilde m^2_X}+\frac{1}{2}\tilde m_X.
\end{equation}

The charge neutral gauge vector boson sector is affected by the extension. Here the $2\times 2$ mass square matrix of the standard model 
is enlarged to become a $3\times 3$ mass square matrix in the $U(1)_X$-extended SUGRA model.
Thus  after spontaneous electroweak symmetry breaking and  the Stueckelberg mass growth the 
$3\times 3$ mass squared matrix of neutral vector bosons in the basis $(C'_{\mu}, B'_{\mu}, A^3_{\mu})$ is given by
\beqn
\mathcal{M}^2_V=\left(\begin{matrix}  M_1^2\kappa^2+\frac{1}{4}g^2_Y v^2 s^2_{\delta} & M_1 M_2\kappa-\frac{1}{4}g^2_Y v^2 s_{\delta} & \frac{1}{4}g_Y g_2 v^2 s_{\delta} \cr
M_1 M_2\kappa-\frac{1}{4}g^2_Y v^2 s_{\delta} & M_2^2+\frac{1}{4}g^2_Y v^2 & -\frac{1}{4}g_Y g_2 v^2 \cr
\frac{1}{4}g_Y g_2 v^2 s_{\delta} & -\frac{1}{4}g_Y g_2 v^2 & \frac{1}{4}g^2_2 v^2 \cr
\end{matrix}\right),
\label{zmassmatrix}
\eeqn
where $A^3_{\mu}$ is the third isospin component, $g_2$ is the $SU(2)_L$ gauge coupling, $\kappa=(c_{\delta}-\epsilon s_{\delta})$, $\epsilon=M_2/M_1$ and $v^2=v^2_u+v^2_d$. The mass-squared matrix of Eq.~(\ref{zmassmatrix}) has one zero eigenvalue which is the photon while the other two eigenvalues are
\begin{align}
&M^2_{\pm} = \frac{1}{2}\Bigg[M_1^2\kappa^2+M^2_2+\frac{1}{4}v^2[g_Y^2 c^2_{\delta}+g_2^2] \nonumber \\
&\pm \sqrt{\left(M_1^2\kappa^2+M^2_2+\frac{1}{4}v^2[g_Y^2 c^2_{\delta}+g_2^2]\right)^2-\Big[M_1^2 g_2^2v^2\kappa^2+M_1^2g^2_Yv^2 c^2_{\delta}+M_2^2g^2_2 v^2\Big]}~\Bigg],
\label{bosons}
\end{align} 
where $M_+$ is identified as the $Z'$ boson mass while $M_-$ as  the $Z$ boson.

\section{Long-lived stau in the $U(1)_X$-extended MSSM/SUGRA}

For this part, we turn off the hidden sector matter as they play no role in the long-lived stau analysis. The $U(1)_X$-extended MSSM/SUGRA model is implemented in the Mathematica package \code{SARAH v4.14.1}~\cite{Staub:2013tta,Staub:2015kfa} which generates model files for \code{SPheno-4.0.3}~\cite{Porod:2003um,Porod:2011nf} which in turn produces the sparticle spectrum and \code{CalcHep/CompHep}~\cite{Pukhov:2004ca,Boos:1994xb} files used by \code{micrOMEGAs-5.0.4}~\cite{Belanger:2014vza} to determine the dark matter relic density and \code{UFO} files which are input to \code{MadGraph5}~\cite{Alwall:2014hca}. The input parameters of the $U(1)_X$-extended MSSM/SUGRA~\cite{msugra} are of the usual non-universal SUGRA model with additional parameters as below (all at the GUT scale)
\begin{equation}
m_0, ~~A_0, ~~ m_1, ~~ m_2, ~~ m_3, ~~M_1, ~~m_X, ~~\delta, ~~\tan\beta, ~~\text{sgn}(\mu),
\label{sugra}
\end{equation}    
where $m_0, ~A_0, ~m_1, ~m_2, ~m_3, ~\tan\beta$ and $\text{sgn}(\mu)$ are the soft parameters in the MSSM sector as defined earlier.  
The parameters $M_2$ and $M_{XY}$ are set to zero at the GUT scale. 
The input parameters must be such as to satisfy a number of experimental constraints. Taking theoretical uncertainties into consideration, we require the Higgs boson mass to be $125\pm 2$ GeV and the dark matter relic density $\Omega h^2 \leq 0.123$. The selected benchmarks satisfying the constraints are shown in Table~\ref{tab1}.          

\begin{table}[H]
\begin{center}
\begin{tabulary}{0.85\textwidth}{l|CCCCCCCCC}
\hline\hline\rule{0pt}{3ex}
Model & $m_0$ & $A_0$ & $m_1$ & $m_2$ & $m_3$ & $M_1$ & $m_X$ & $\tan\beta$ & $\delta$ \\
\hline\rule{0pt}{3ex}  
\!\!(a)& 300 & 1838 & 885 & 740 & 4235 & 473 & 600 & 14 & $2.0\times10^{-5}$ \\
(b)  & 546 & -3733 & 828 & 761 & 3657 & 426 & 392 & 16 & $4.7\times10^{-6}$ \\
(c)  & 529 & -3211 & 864 & 482 & 3777 & 461 & 400 & 15 & $6.0\times10^{-6}$ \\
(d)  & 680 & -5198 & 1166 & 806 & 3945 & 503 & 198 & 15 & $2.5\times10^{-6}$ \\
(e)  & 563 & -1850 & 1214 & 598 & 3856 & 579 & 380 & 21 & $2.4\times10^{-6}$ \\
(f)  & 500 & -2698 & 1286 & 893 & 4165 & 523 & 65 & 15 & $2.5\times10^{-6}$ \\
(g)  & 515 & -261 & 1451 & 1265 & 4830 & 682 & 258 & 25 & $1.4\times10^{-6}$ \\
(h)  & 645 & 1009 & 1621 & 1160 & 5374 & 714 & 100 & 26 & $1.3\times10^{-6}$ \\
\hline
\end{tabulary}\end{center}
\caption{Input parameters for the benchmarks used in the analysis of Ref.~\cite{Aboubrahim:2019qpc}. Here $M_2=M_{XY}=0$ at the GUT scale. All masses are in GeV.}
\label{tab1}
\end{table}

The resulting spectrum of some of the relevant particles is shown in Table~\ref{tab2}. 

\begin{table}[H]
\begin{center}
\begin{tabulary}{1.3\textwidth}{l|CCCCCCCCCCC}
\hline\hline\rule{0pt}{3ex}
Model  & $h^0$ & $\mu$ & $\tilde\chi_1^0$ & $\tilde\chi_1^\pm$ & $\tilde{\tau}$ & $\tilde{\nu}_{\tau}$ & $\tilde{\xi}^0_1$ & $\tilde t$ & $\tilde g$ & $\Omega h^2$ & $c\tau_0$ \\
\hline\rule{0pt}{3ex} 
\!\!(a) & 123.0 & 4127 & 359.9 & 556.9 & 275.1 & 434.3 & 260.1 & 6306 & 8459 & 0.116 & 243.6 \\
(b) & 123.1 & 4417 & 343.3 & 595.2 & 291.0 & 572.4 & 272.9 & 5118 & 7372 & 0.123 & 199.9 \\
(c) & 123.4 & 4426 & 350.3 & 350.5 & 319.3 & 459.8 & 302.5 & 5376 & 7621 & 0.109 & 147.0 \\
(d) & 124.6 & 4998 & 495.2 & 633.2 & 428.0 & 671.4 & 413.6 & 5347 & 7916 & 0.121 & 177.6 \\
(e) & 123.1 & 4236 & 449.0 & 449.2 & 440.5 & 570.6 & 419.4 & 5607 & 7764 & 0.111 & 307.6 \\
(f) & 124.2 & 4669 & 546.0 & 699.7 & 500.0 & 653.6 & 491.5 & 5926 & 8326 & 0.119 & 387.3 \\
(g) & 123.2 & 4852 & 619.4 & 1009 & 583.0 & 864.7 & 565.1 & 6997 & 9553 & 0.114 & 424.1 \\
(h) & 123.4 & 5193 & 692.8 & 911.3 & 680.8 & 877.3 & 665.7 & 7816 & 10572 & 0.120 & 561.3 \\
\hline
\end{tabulary}\end{center}
\caption{Display of the Higgs boson ($h^0$) mass, the $\mu$ parameter, the stau mass,  the relevant electroweak gaugino masses, and the relic density for the benchmarks  of Table~\ref{tab1} computed at the electroweak scale (taken from Ref.~\cite{Aboubrahim:2019qpc}). The track length, $c\tau_0$ (in mm) left by the long-lived stau is also shown. All masses are in GeV.}
\label{tab2}
\end{table}

All sparticles shown in Table~\ref{tab2} have masses which are not ruled out by experiment yet. Here the stau is the lighter of the two staus and considering large off-diagonal element in the stau mass-squared matrix it can be made lighter than the tau sneutrino. The mass gap between the NLSP and the hidden sector LSP is small which contributes to the stau decay width suppression.  The only decay mode of the stau is to the hidden sector neutralino, i.e. $\tilde\tau \rightarrow \tau\tilde\xi^0_1$. Another source of suppression comes from the fact that the MSSM particles communicate with the hidden sector particles only through the small kinetic mixing coefficient $\delta$ which, according to Table~\ref{tab1}, is chosen to be very small, i.e. $\mathcal{O}(10^{-6})$.

\subsection{LHC production of stau and its long-lived decay signature}

At the LHC, the light stau can be pair produced or produced in association with a tau sneutrino, i.e. $pp\rightarrow \tilde\tau^{+}\tilde\tau^{-}$ and $pp\rightarrow \tilde\tau\tilde\nu_{\tau}$, respectively. The stau production in the extended model proceeds via $s$-channel $\gamma$, $Z$ and $Z'$ mediators with the contribution of the latter being negligible due to the smallness of the gauge kinetic and mass mixings whereas stau associated production with a tau sneutrino proceeds by the exchange of $W^{\pm}$ boson. We calculate the di-stau and stau-tau sneutrino LHC production cross-sections using \code{Prospino2}~\cite{Beenakker:1996ed,Beenakker:1999xh} at the next-to-leading order (NLO) in QCD at 14 TeV and at 27 TeV using the CTEQ5 PDF set~\cite{Lai:1999wy}. The cross-sections of the selected benchmarks can be found in Ref.~\cite{Aboubrahim:2019qpc}.   

The signal final states following the production of staus and tau sneutrinos are as in Eq.~(\ref{finalstates})
\begin{align}
pp\rightarrow \tilde\tau^+\tilde\tau^-\rightarrow \tau^+\tau^-\tilde\xi^0_1\tilde\xi^0_1\rightarrow \tau_h,\ell+E^{\rm miss}_T, \nonumber  \\
pp\rightarrow \tilde\tau^{\pm}\tilde\nu_{\tau}\rightarrow \tau^{\pm}\tilde\xi^0_1\tau^{\pm}W^{\mp}\rightarrow \tau_h,2\ell+E^{\rm miss}_T,
\label{finalstates}
\end{align}
where $\tau_h$ corresponds to a hadronically decaying $\tau$, $\ell$ represents a light lepton (electron or muon) and $E^{\rm miss}_T$ is the missing transverse energy due to neutrinos and the LSP. As Eq.~(\ref{finalstates}) suggests, we require at least one isolated light lepton and at most one hadronically decaying tau. Due to the compressed spectrum, little $E^{\rm miss}_T$ is produced and so no selection criteria is imposed on the missing transverse energy. 

There are two sets of tracks that may be visible in the ID tracker: one corresponding to the long-lived stau which is characterized by low speed and large invariant mass and the second due to soft leptons (of low $p_T$) coming from the stau decay. The combination of both tracks constitute what is known as a kinked track~\cite{Lee:2018pag}. The main selection criterion involves cuts on the lepton impact parameter $d_0$ which tends to be much larger than the SM backgrounds. Given the final states of Eq.~(\ref{finalstates}), the largest contributors to the physical SM backgrounds are $W/Z/\gamma^*+$ jets, diboson production, single top and $t\bar{t}$. The signal and background events are simulated at leading order (LO) with \code{MadGraph5\_aMC@NLO-2.6.3} interfaced to \code{LHAPDF}~\cite{Buckley:2014ana}, showered by \code{PYTHIA8}~\cite{Sjostrand:2014zea} and detector simulation and event reconstruction carried out by \code{DELPHES-3.4.2}~\cite{deFavereau:2013fsa} using the beta card for HL-LHC and HE-LHC studies. \code{ROOT 6}~\cite{Antcheva:2011zz} is used for the analysis of the resulting event files and cut implementation. The unphysical background contamination is due to fake tracks arising from the high pile-up environment. We simulate minimum bias events due to elastic and inelastic (diffractive and non-diffractive) soft QCD events with \code{PYTHIA8} which are mixed with the main interaction. We consider a mean pile-up (interactions per bunch crossing) of 128~\cite{ATLAS:2013hta} for both HL-LHC and HE-LHC\footnote{Estimated pile-up at the HL-LHC may reach $\sim 200$ while at HE-LHC the figure may rise up to $\sim 800$.}. Pile-up mitigation is handled by \code{PUPPI}~\cite{Bertolini:2014bba} with the default settings used for CMS phase II Delphes card.

\subsection{Discovery potential of long-lived stau at the LHC}

In order to see the extent of performance degradation due to pile-up we carry out the analysis in cases of zero and non-zero pile-up. As explained before, we are looking for a light lepton (electron or muon) track with large impact parameter, $|d_0|$, originating from a high momentum track due to the long-lived charged stau. The kinematic variables used to discriminate the signal from the SM background are:
\begin{enumerate}
\item $|d_0|$: the track impact parameter which is chosen to be large enough to eliminate as many background events as possible.
\item $p_T^{e~[\mu]}$: the transverse momentum of an isolated electron or muon. 
\item $p_T^{\rm tracks}$: the transverse momentum of tracks in the ID.
\item Isolated lepton tracks: the number of isolated leptons must match the number of lepton tracks. This ensures that we reject any lepton tracks which are not isolated. 
\item $\Delta R(\tilde\tau,\rm track)$: the minimum spatial separation between the lepton tracks and the stau track. A small cut on this variable ensures that the lepton track considered has originated from a long-lived stau.
\item $\beta=p/E$: the velocity of the long-lived particle. A cut on $\beta$ allows us to reject events with muons faking a stau track.   
\end{enumerate}

\begin{figure}[H]
\centering
\includegraphics[width=0.49\textwidth]{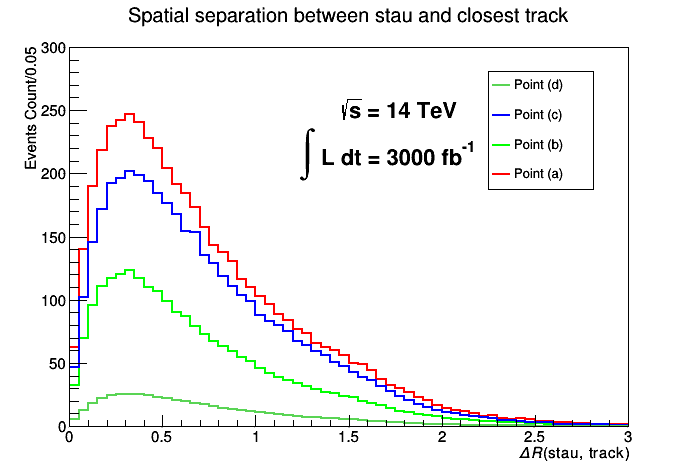}
\includegraphics[width=0.49\textwidth]{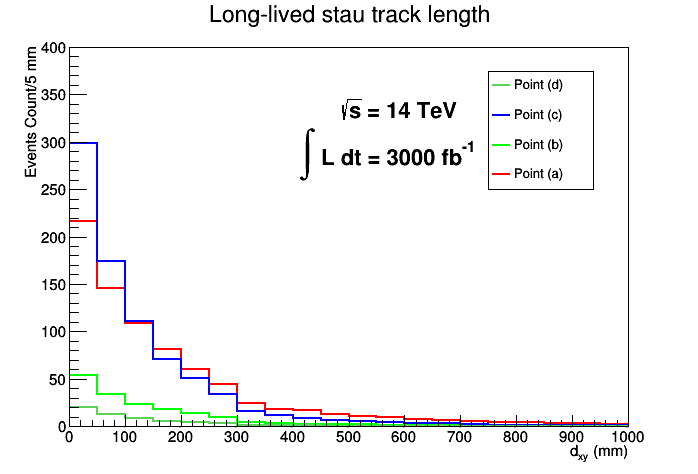}
\caption{Left panel: Minimum spatial separation between the stau LLP and its closest lepton track, $\Delta R(\tilde\tau,\rm track)$. Right panel: the track length $d_{xy}$, of the long-lived stau~\cite{Aboubrahim:2019qpc}.}
\label{fig1}
\end{figure}

We present in the left panel of Fig.~\ref{fig1} $\Delta R(\tilde\tau,\rm track)$ which has peak values for small spatial separation. Thus a cut of $\Delta R(\tilde\tau,\rm track)<0.6$ should be sufficient to ensure that the lepton tracks have actually originated from the corresponding stau track. The right panel of Fig.~\ref{fig1} displays the decay length, $d_{xy}$ of the stau which clearly can travel up to 1 m in the ID, knowing that the typical tracker radius is between 35 mm and 1200 mm.   

\begin{figure}[H]
\centering
\includegraphics[width=0.48\textwidth]{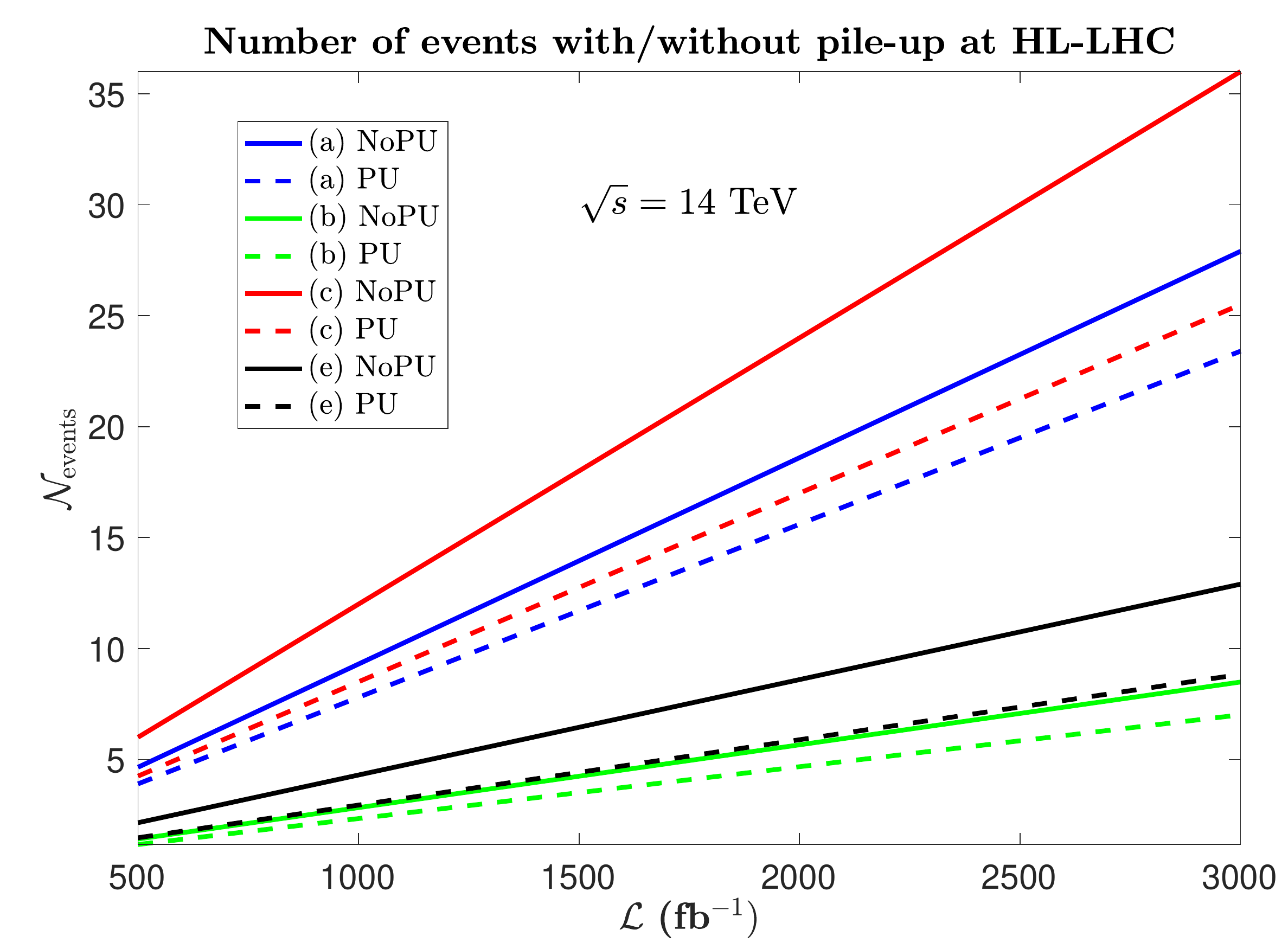}
\includegraphics[width=0.48\textwidth]{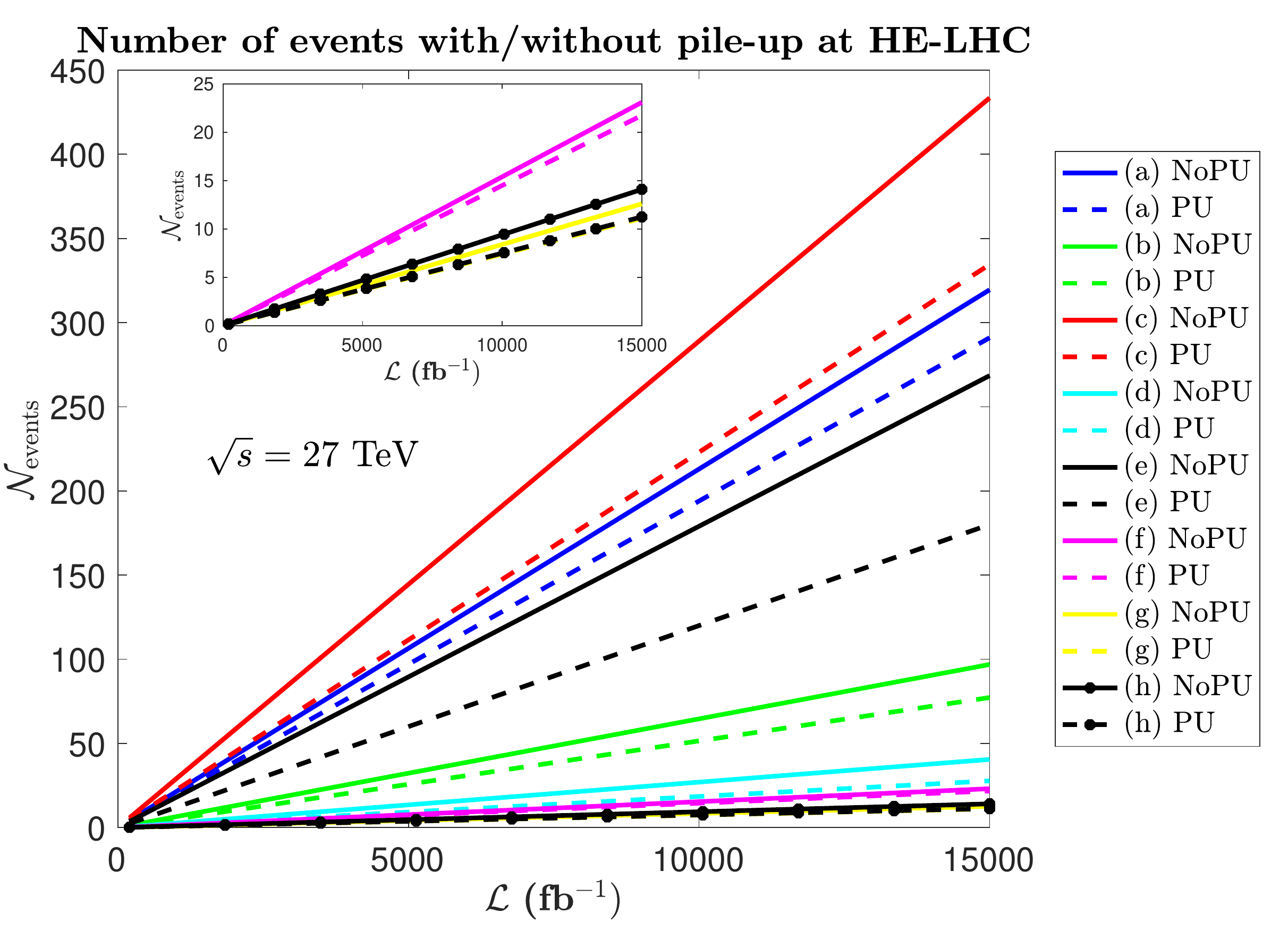}
\caption{Left panel: Estimated number of events for various integrated luminosities for benchmarks (a), (b), (c) and (e) in cases of no pile-up (solid lines) and pile-up (dashed lines) at HL-LHC. Right panel: same as the left panel but for HE-LHC for all the benchmarks of Table~\ref{tab1} (taken from Ref.~\cite{Aboubrahim:2019qpc}).}
\label{fig2}
\end{figure}

Applying some selection criteria on the kinematic variables listed above, we perform a cut-and-count analysis to determine the integrated luminosity needed for a $5\sigma$ discovery at both HL-LHC and HE-LHC under zero and non-zero pile-up conditions. The number of signal events surviving the cuts for cases of pile-up and no pile-up are displayed in Fig.~\ref{fig2} at 14 TeV and at 27 TeV as a function of the integrated luminosity. For the 14 TeV case, only the observable points are displayed (all points are observable in the 27 TeV case). It is seen that the event yield has dropped by amounts ranging from $\sim 16\%$ for point (a) to $\sim 30\%$ for point (e) at 14 TeV and from $\sim 6\%$ for point (f) to $\sim 32\%$ for point (e) at 27 TeV. 

\section{Multicomponent DM model in the $U(1)_X$-extended MSSM/SUGRA}\label{sec:benchmarks}

For this part of the analysis we include the hidden sector matter. The $6\times 6$ neutralino mass matrix is now written in the rotated basis $(\lambda'_Y,\lambda_3,\tilde h_1, \tilde h_2,\lambda'_X,\psi_S)$ and whose mass eigenstates are 
\begin{equation}
 \tilde \chi_1^0, ~\tilde \chi_2^0, ~\tilde \chi_3^0, ~\tilde \chi_4^0, ~\tilde \chi_5^0, ~\tilde \chi_6^0\, ,
 \end{equation}
where $\tilde \chi_5^0$ and $\tilde \chi_6^0$ belong to the hidden sector. The particle content of the $U(1)_X$-extended MSSM/SUGRA consists of the particles of the MSSM, and from the hidden sector three spin 1/2 particles (a Dirac fermion $\psi$ and two Majorana neutralinos $\tilde\chi^0_5$, $\tilde\chi^0_6$), three spin zero particles ($\rho$, $\phi$, $\phi'$) and one massive  vector boson $Z'$. The input parameters of the $U(1)_X$-extended MSSM/SUGRA with hidden sector matter are taken to be 
$$m_0, ~~A_0, ~~ m_1, ~~ m_2, ~~ m_3, ~~M_1, ~~m_X, ~~M_{\psi}, ~~B_{\psi}, ~~\delta, ~~g_X, ~~\tan\beta, ~~\text{sgn}(\mu),$$    
where $B_{\psi}$ is the bilinear parameter of the Dirac mass term in the superpotential at the GUT scale. 

We show in Table~\ref{tab3} ten representative benchmarks satisfying the Higgs boson mass and relic density constraints. The values of $\mu$ are $\mathcal{O}(100)$ GeV which support arguments for naturalness. Since the mass of the CP odd Higgs depends on $\mu B$ one expects light CP odd states. 

\begin{table}[H]
\begin{center}
\resizebox{1.05\textwidth}{!}{\begin{tabulary}{\textwidth}{l|ccccccccccccc}
\hline\hline\rule{0pt}{3ex}
Model & $m_0$ & $A_0$ & $m_1$ & $m_2$ & $m_3$ & $\mu$ & $M_1$ & $m_X$ & $M_{\psi}$ & $B_{\psi}$ & $\tan\beta$ & $g_X$ & $\delta$ \\
\hline\rule{0pt}{3ex}  
\!\!(a) & 8115 & -7477 & 6785 & 9115 & 4021 & 423 & 1261 & 27 & 627 & 9283 & 6 & 0.06 & 0.02 \\
(b) & 1743 & 898 & 4551 & 2160 & 4084 & 301 & -1086 & 27 & 627 & 5167 & 10 & 0.07 & 0.02 \\
(c) & 1056 & -920 & 1706 & 3417 & 3396 & 243 & 1059 & 89 & 525 & 2846 & 10 & 0.03 & 0.01 \\
(d) & 8424 & -2488 & 6165 & 3544 & 2466 & 330 & -1469 & 473 & 733 & 4680 & 12 & 0.03 & 0.01 \\
(e) & 2011 & -2462 & 3008 & 5030 & 3833 & 598 & 875 & 38 & 425 & 3248 & 9 & 0.06 & 0.06 \\
(f) & 4637 & -4045 & 7004 & 5480 & 2727 & 511 & -1230 & 372 & 613 & 7557 & 15 & 0.04 & 0.04 \\
(g) & 819 & 477 & 7847 & 1218 & 3040 & 201 & 820 & 509 & 401 & 3425 & 12 & 0.05 & 0.09 \\
(h) & 3881 & -2580 & 7449 & 4870 & 4429 & 268 & 850 & 152 & 419 & 9199 & 13 & 0.08 & 0.02 \\
(i) & 1349 & -2722 & 3938 & 4420 & 2558 & 482 & 1292 & 19 & 636 & 4235 & 15 & 0.07 & 0.08 \\
(j) & 2015 & -4435 & 2695 & 5399 & 2470 & 217 & 1343 & 690 & 670 & 4587 & 11 & 0.03 & 0.03 \\ 
\hline
\end{tabulary}}\end{center}
\caption{Input parameters for the benchmarks used in the analysis of Ref.~\cite{Aboubrahim:2019vjl}. Here $M_{XY}=0=B$ at the GUT scale and  $M_2$ is chosen at the GUT scale so that it is nearly vanishing at the electroweak scale. All masses are in GeV.}
\label{tab3}
\end{table}

The CP odd Higgs mass ranging from $\sim 300$ GeV to 750 GeV along with the neutralino, chargino, stop, gluino and stau masses are presented in Table~\ref{tab4}. In some of those benchmarks, the value of $m_0$ is quite small, for instance, point (g) has $m_0 \sim 800$ GeV while the stop mass is $\sim 5$ TeV. The reason is the large value of $m_3$ which via the RGE running generates squark masses in the several TeV range~\cite{Akula:2013ioa}. With heavy gluinos and stops, experimental limits on their masses from ATLAS and CMS can be evaded. Also, the LSP and chargino masses presented in Table~\ref{tab4} have not yet  been ruled out by experiment. Note that higgsino-like neutralinos contribute the smaller fraction of the DM relic density (due to the smallness of $\mu$) while the Dirac fermion of the hidden sector contributes the larger fraction. 

\begin{table}[H]
\begin{center}
\begin{tabulary}{1.3\textwidth}{l|CCCCCCCCCCC}
\hline\hline\rule{0pt}{3ex}
Model  & $h$ & $\tilde\chi_1^0$ & $\tilde\chi_1^\pm$ & $\tilde{\tau}$ & $\tilde{\chi}^0_5$ & $\tilde t$ & $\tilde g$ & $A$ & $\Omega h^2$ & $(\Omega h^2)_{\chi}$ & $(\Omega h^2)_{\psi}$ \\
\hline\rule{0pt}{3ex} 
\!\!(a) & 123.3 & 455.9 & 457.1 & 8109 & 1245 & 6343 & 8408 & 305.8 & 0.124 & 0.022 & 0.102 \\
(b) & 123.3 & 322.6 & 324.9 & 2115 & 1008 & 5898 & 8195 & 351.8 & 0.101 & 0.012 & 0.089 \\
(c) & 123.1 & 258.9 & 262.6 & 665.6 & 1015 & 4565 & 6855 & 408.9 & 0.116 & 0.009 & 0.107 \\
(d) & 124.0 & 354.8 & 356.4 & 8425 & 1250 & 6573 & 5467 & 450.8 & 0.117 & 0.019 & 0.098 \\
(e) & 123.9 & 639.5 & 642.2 & 1875 & 851.5 & 4943 & 7712 & 504.2 & 0.106 & 0.042 & 0.064 \\
(f) & 124.7 & 544.3 & 545.7 & 4982 & 1055 & 4314 & 5803 & 547.3 & 0.125 & 0.031 & 0.094 \\
(g) & 123.1 & 212.4 & 215.3 & 1906 & 601.8 & 4646 & 6229 & 604.2 & 0.118 & 0.006 & 0.112 \\
(h) & 125.0 & 289.1 & 290.5 & 4426 & 775.5 & 6109 & 8565 & 650.9 & 0.121 & 0.009 & 0.112 \\
(i) & 124.3 & 510.8 & 512.9 & 1627 & 1276 & 3077 & 5292 & 702.7 & 0.118 & 0.028 & 0.090 \\
(j) & 125.0 & 231.5 & 233.7 & 1845 & 1041 & 2335 & 5164 & 750.3 & 0.113 & 0.008 & 0.105 \\
\hline
\end{tabulary}\end{center}
\caption{Display of the SM-like Higgs boson mass, the stau mass,  the relevant electroweak gaugino masses, the CP odd Higgs mass and the relic density for the benchmarks of Table~\ref{tab3} computed at the electroweak scale (taken from Ref.~\cite{Aboubrahim:2019vjl}). All masses are in GeV. }
\label{tab4}
\end{table}

Corrections to the $Z$ boson mass due to gauge kinetic and mass mixings are well within experimental error bars. Additionally, the particle spectrum of the model contains an extra neutral massive gauge boson, $Z'$. Stringent constraints are set on the mass of the $Z'$~\cite{Tanabashi:2018oca} and most recently by ATLAS~\cite{Aad:2019fac} using 139$\ifb$ of data. In new physics models containing $Z'$ with SM couplings, the mass limit is set at $m_{Z'}>5.1$ TeV. For a model with an extra $U(1)_X$ with a gauge coupling strength $g_X$, the limit can be written as 
\begin{equation}
\frac{m_{Z'}}{g_X}\gtrsim 12 ~\text{TeV},
\label{zlimit}
\end{equation}
which is satisfied for all the benchmarks of Table~\ref{tab3}.

\subsection{Associated production of CP odd Higgs  with heavy quarks}
\label{sec:lhcproduction}

As a result of electroweak symmetry breaking and the Stueckelberg mass growth, the Higgs sector of the $U(1)_X$-extended MSSM has six degrees of freedom corresponding to three CP even Higgs, $h, H$ and $\rho$ and one CP odd Higgs $A$ and two charged Higgs $H^{\pm}$. In this section we discuss the production and decay of the CP odd Higgs $A$ and in particular the associated production of $A$ with bottom-anti bottom quarks, $b\bar{b}A$. 

There are two approaches to calculating the production cross-section of $A$ in association with $b\bar{b}$. The first is known as the four-flavour scheme (4FS) which considers the $b$-quark to be heavy and appearing only in the final state. The LO partonic processes are
\begin{equation}
gg\rightarrow b\bar{b}A, ~~~ q\bar{q}\rightarrow b\bar{b}A. 
\end{equation}

The cross-section appears to be sensitive to the $b$-quark mass which is taken to be the running mass at the appropriate renormalization and factorization scales ($\bar{m}_b(\mu_F)$). As mentioned before, the CP odd Higgs coupling to the top quark is suppressed and thus the diagrams involving top quarks do not contribute significantly to the cross-section. Following the prescription of~\cite{deFlorian:2016spz}, the hard scale of the process is taken at the renormalization and factorization scales such that $\mu_R=\mu_F=(m_A+2m_b)/4$ with $m_A$ the CP odd Higgs mass and $m_b$ being the $b$-quark pole mass. The 4FS NLO cross-section at fixed order in $\alpha_S$ is calculated with \code{MadGraph5\_aMC@NLO-2.6.3} using \code{FeynRules}~\cite{Alloul:2013bka} \code{UFO} files~\cite{Degrande:2011ua,Degrande:2014vpa} for the Type-II two Higgs doublet model (2HDM). The choice of the latter is justified due to the fact that SUSY-QCD effects for our benchmarks are very minimal since the squarks and gluinos and heavy. The cross-sections at 14 TeV and 27 TeV are given in Ref.~\cite{Aboubrahim:2019vjl} and have been checked with publically available results~\cite{deFlorian:2016spz} by a proper scaling of the bottom Yukawa coupling. 

The second method to calculating the cross-section is known as the five-flavour scheme (5FS) which is realized when divergent logarithmic terms are absorbed to all orders in $\alpha_S$ via the DGLAP evolution of $b$-quark PDFs. In this scheme, $b$-quarks are massless and appear at the partonic level so the LO process for CP odd Higgs production is
\begin{equation}
b\bar{b}\rightarrow A.
\end{equation}

The 5FS $b\bar{b}A$ production cross-section is known at next-to-NLO (NNLO) and here we use \code{SusHi-1.7.0}~\cite{Harlander:2012pb} to determine those cross-sections at 14 TeV and 27 TeV. Suitable choices of the renormalization and factorization scales are $\mu_R=m_A$ and $\mu_F=m_A/4$, respectively. Scale uncertainties are determined by varying $\mu_R$ and $\mu_F$ such that $\mu_R, 4\mu_F\in\{m_A/2,m_A,2m_A\}$ with $1/2\leq 4\mu_F/\mu_R<2$. Although the $b$-quark is massless, the bottom Yukawa coupling is non-zero and renormalized in the $\overline{\rm MS}$ scheme. In calculating the cross-sections for both 4FS and 5FS cases we have used \code{PDF4LHC15\_nlo\_mc} and \code{PDF4LHC15\_nnlo\_mc}~\cite{Butterworth:2015oua} PDFs, respectively. We combine both estimates of the cross-section using the Santander matching criterion~\cite{Harlander:2011aa} such that 
\begin{equation}
\sigma^{\rm matched}=\frac{\sigma^{4\rm FS}+\alpha\sigma^{5\rm FS}}{1+\alpha},
\label{matched}
\end{equation}  
where $\alpha=\ln\left(\frac{m_A}{m_b}\right)-2$. The matched cross-section of the inclusive process lies between the 4FS and 5FS values but closer to the 5FS value owing to the weight $\alpha$ which depends on the CP odd Higgs mass. The uncertainties are combined as such, 
\begin{equation}
\delta\sigma^{\rm matched}=\frac{\delta\sigma^{4\rm FS}+\alpha\delta\sigma^{5\rm FS}}{1+\alpha}.
\end{equation}

\subsection{CP odd Higgs signature in $\tau_h\tau_h$ final state at the LHC}\label{sec:signature}

Due to its enhanced coupling to bottom quarks, the CP odd Higgs preferentially decays to $b\bar{b}$ pair while the second largest branching ratio is to $\tau^+\tau^-$. The benchmarks of Table~\ref{tab3} are not yet excluded by experiment and lie within the contour set for HL-LHC. 

The signal we investigate consists of a CP odd Higgs decaying to two hadronic taus and produced alongside two $b$-quarks which can be tagged. Even in the 5FS, $b$-flavored jets can appear at the parton shower level and so $b$-tagging is viable here too. In order to account for misidentified $b$-tagged jets, we require that our final states contain at least one $b$-tagged jet and two tau-tagged $(\tau_h)$ jets such that $p_T(b)>20$ GeV, $|\eta(b)|<2.5$ and $p_T(\tau_h)>15$ GeV. The standard model backgrounds relevant to the final states considered here are $t\bar{t}$, $t$+jets, $t+W/Z$, QCD multijet, diboson and $W,Z/\gamma^*$+jets. 

Due to the smallness of the signal cross-section in comparison to the SM backgrounds (especially following the selection criteria), we use Boosted Decision Trees (BDT) to separate the signal from the background. The type of BDT used here is known as ``Adaptive BDT" (\code{AdaBoost}). The BDT implementation is carried out using \code{ROOT}'s own TMVA (Toolkit for Multivariate Analysis) framework~\cite{Speckmayer:2010zz}. Depending on the samples, we set the number of trees to be in the range 120 to 200, the depth to 3 and the \code{AdaBoost} learning rate to 0.5. The kinematic variables used in discriminating the signal from the background are: the total transverse mass of the di-tau system given by~\cite{ATLAS:2016fpj}
\begin{equation}
m^{\rm tot}_T=\sqrt{m^2_T(E^{\rm miss}_T,\tau_{h1})+m^2_T(E^{\rm miss}_T,\tau_{h2})+m^2_T(\tau_{h1},\tau_{h2})},
\end{equation}
where
\begin{equation}
m_T(i,j)=\sqrt{2p_T^i p_T^j(1-\cos\Delta\phi_{ij})},
\end{equation}
is the hadronic di-tau invariant mass, $m_{\tau_h\tau_h}$, the angular separation $\Delta\phi(\tau_{h1},\tau_{h2})$ between the leading and sub-leading hadronic tau jets, the number of charged tracks associated with the leading tau, $N_{\rm tracks}^{\tau}$, the number of $b$-tagged jets $N_{\rm jet}^b$, the variable $\ln(p_T^{\rm jet})$ defined as
\begin{equation}
\ln(p_T^{\rm jet})= \begin{cases} 
      \ln(p_T^{\rm jet_1}) & \text{if} ~~N_{\text{jets}}\geq 1 \\
      0 & \text{if} ~~N_{\text{jets}}=0 
   \end{cases}~~,
\end{equation}
where $p_T^{\rm jet_1}$ is the $p_T$ of the leading jet, the di-jet transverse mass $m^{\rm di-jet}_T$ of the leading and sub-leading jets and the effective mass defined as
\begin{equation}
m_{\rm eff}=H_T+E^{\rm miss}_T+p_T(\tau_{h1})+p_T(\tau_{h2}),
\end{equation}
where $H_T$ is the sum of the hadronic $p_T$'s in an event, $p_T(\tau_{h1})$ and $p_T(\tau_{h2})$ are the transverse momenta of the leading and sub-leading hadronic taus. 

After training and testing of the BDTs, we set a cut on the BDT score variable which would give us the minimum integrated luminosity for $\frac{S}{\sqrt{S+B}}$ at the $5\sigma$ level discovery. The BDT distribution varies from one benchmark to another due to the fact that each point is trained along with the SM backgrounds independently from the others. For this reason, the most efficient cut on the BDT score will not be the same for all points. We present in Fig.~\ref{fig4} the computed 
 integrated luminosities, $\mathcal{L}$, as a function of the cut on the BDT score for both 14 TeV (left panel) and 27 TeV (right panel) machines. For 14 TeV, one can see that a drop in $\mathcal{L}$ occurs for BDT score $>0.3$ while at 27 TeV the same is seen for BDT score $>0.2$.   

\begin{figure}[H]
 \centering
 	\includegraphics[width=0.48\textwidth]{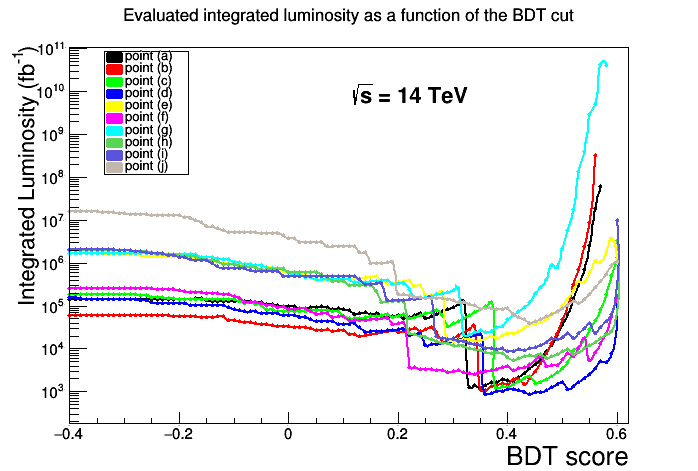}
 	\includegraphics[width=0.48\textwidth]{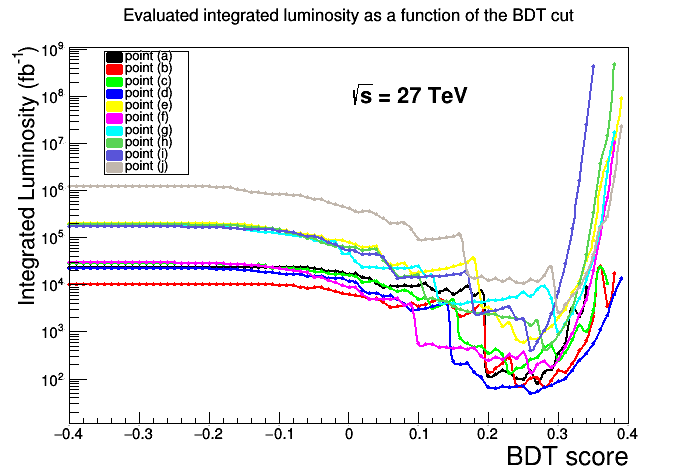} 
      \caption{The estimated      
       integrated luminosities as a function of the BDT cut for the benchmarks of Table~\ref{tab3} at 14 TeV (left panel) and 27 TeV (right panel) from Ref.~\cite{Aboubrahim:2019vjl}.}
	\label{fig4}
\end{figure}

It is shown that half the benchmarks are discoverable at HL-LHC with benchmark (d) discoverable with an $\mathcal{L}$ as low as 866$\ifb$ while the benchmark (f)  requires  $\mathcal{L}$ close to the optimal integrated luminosity of 3000$\ifb$. However, all  the benchmarks are discoverable at HE-LHC with some requiring an integrated luminosity smaller than 100$\ifb$ 
 such as point (d) with $\mathcal{L}=50\ifb$ for discovery. Point (j) requires the largest amount of data at $\sim 2600\ifb$ 
 which, however,  is still much lower than the optimal integrated luminosity of 15 ab$^{-1}$ expected at HE-LHC.

We show in Fig.~\ref{fig5} distributions in the BDT score for points (a) and (d) at 14 TeV and 27 TeV for some select integrated luminosities. For point (a) which is observable at both HL-LHC and HE-LHC we  see that the signal is over the background for $\mathcal{L}=150 \ifb$ at 27 TeV (top left panel) while higher integrated luminosity is required for an excess at HL-LHC, namely,  $\mathcal{L}=2000 \ifb$ (top right panel). The bottom two panels of Fig.~\ref{fig5} show the BDT score for point (d) at 27 TeV (left) and 14 TeV (right) for the same integrated luminosity of $200\ifb$. 

\begin{figure}[H]
 \centering
 	\includegraphics[width=0.45\textwidth]{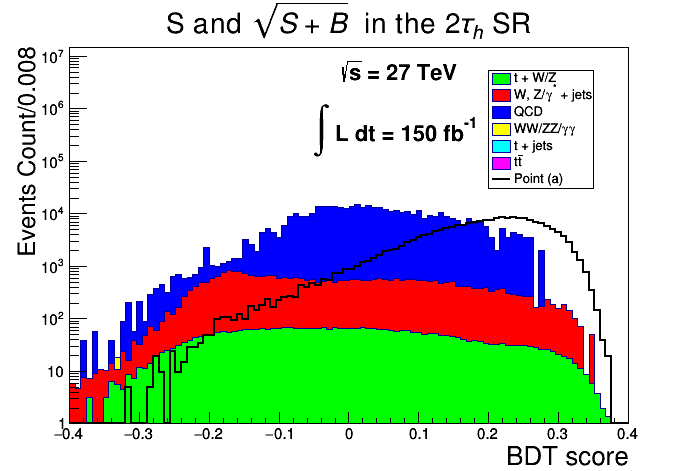}
 	\includegraphics[width=0.45\textwidth]{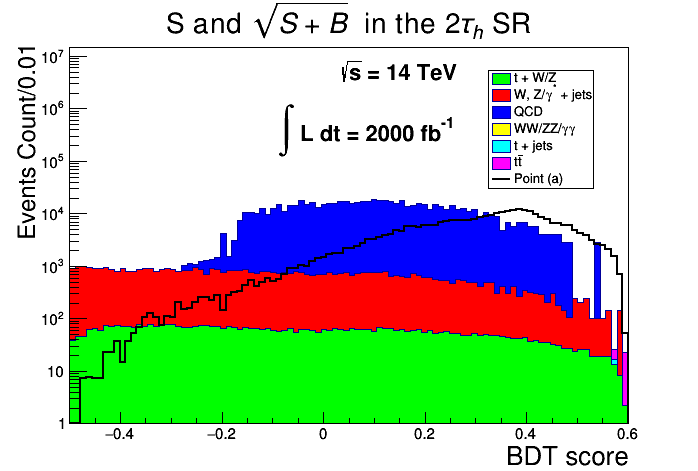}\\
 	\includegraphics[width=0.45\textwidth]{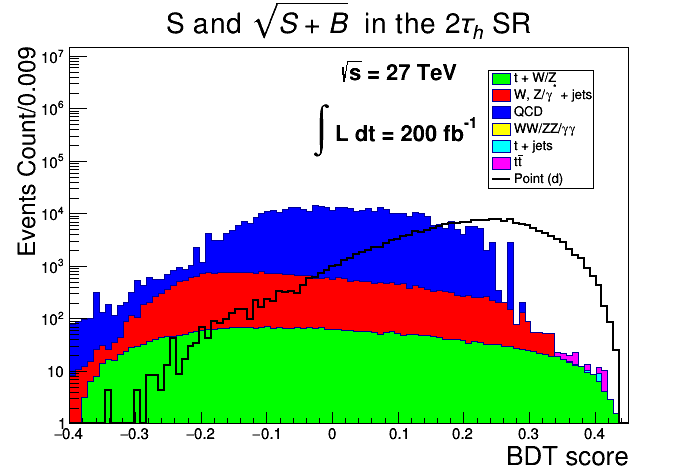}
 	\includegraphics[width=0.45\textwidth]{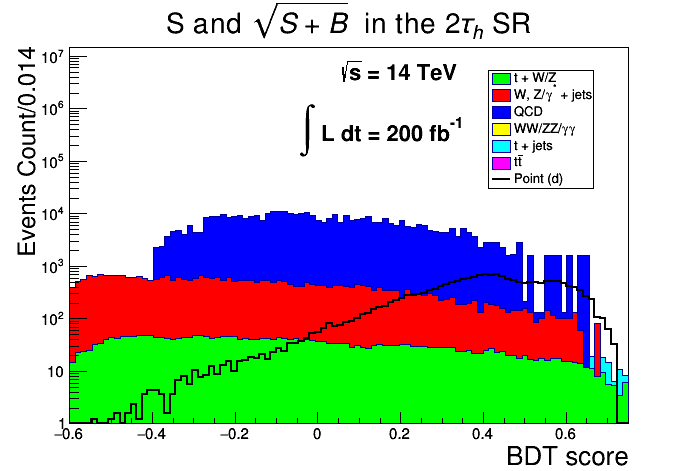}
      \caption{Distributions in the BDT score for benchmarks (a) (top panels) and (d) (bottom panels) of Table~\ref{tab3} at 14 TeV (right panels) and 27 TeV (left panels) in the $2\tau_h$ signal region (SR) for various integrated luminosities obtained from Ref.~\cite{Aboubrahim:2019vjl}. }
	\label{fig5}
\end{figure}

The HE-LHC is projected to collect data at a rate of 820$\ifb$ per year~\cite{HE-LHC-1} which is much higher than the HL-LHC rate. Hence the expected run time for points (a)$-$(d) for discovery at HL-LHC is $\sim 3$ to $\sim 4$ years while point (f) requires $\sim 8$ years. The run time is significantly decreased for HE-LHC where most of the points require $\sim 1$ to $\sim 6$ months, while point (d) $\sim 22$ days and point (j) $\sim 3$ years.   

Before concluding we give an overview of the uncertainties one might expect and their impact on the estimated integrated luminosities at HL-LHC and HE-LHC. Studies show that much of the systematic uncertainties are expected to drop by a factor of 2~\cite{Cepeda:2019klc,CidVidal:2018eel} and now known as ``YR18 systematic uncertainties". We exhibit in Fig.~\ref{fig6} the estimated integrated luminosities before and after including the uncertainties for both HL-LHC and HE-LHC. In the left panel, the five benchmarks discoverable at both colliders are shown along with the ``YR18 uncertainties" where at HL-LHC, the integrated luminosities for discovery have increased by $\sim 1.5$  to $\sim 2.5$ times (in blue) compared to when no systematic uncertainties are present (in orange). At the HE-LHC the increase is by $\sim 1.5$ to $\sim 4$ times (in red) compared to the case with no systematics (in yellow). The right panel shows the points that are discoverable only at HE-LHC along with the integrated luminosities before (in orange) and after (in blue) including uncertainties. In the analysis above we have not included any CP phases. Such phases are known to have measurable effects  in SUSY and Higgs phenomena (see e.g., \cite{CPphases}) and is of interest in future work.

\begin{figure}[H]
 \centering
 	\includegraphics[width=0.49\textwidth]{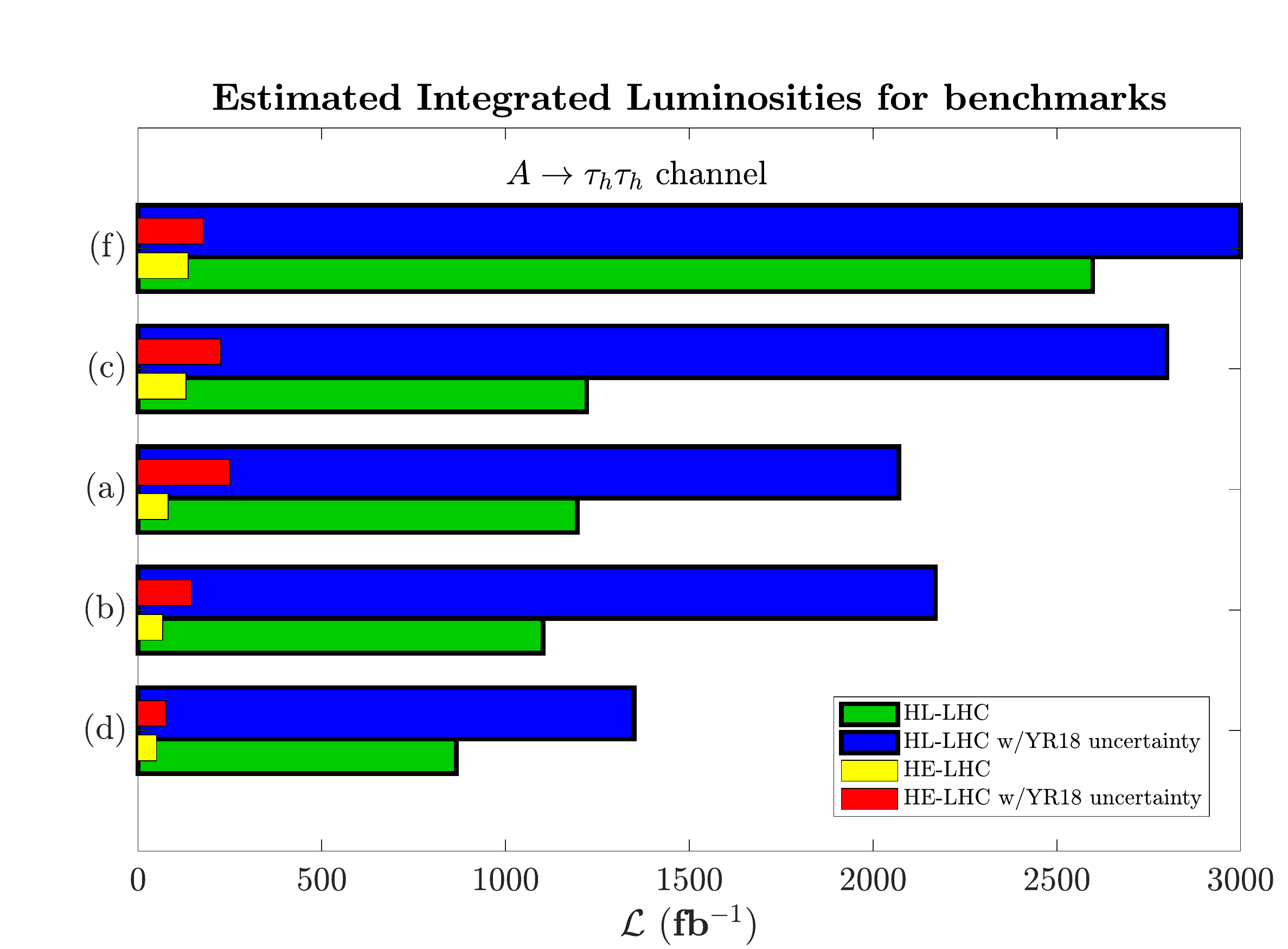}
 	\includegraphics[width=0.49\textwidth]{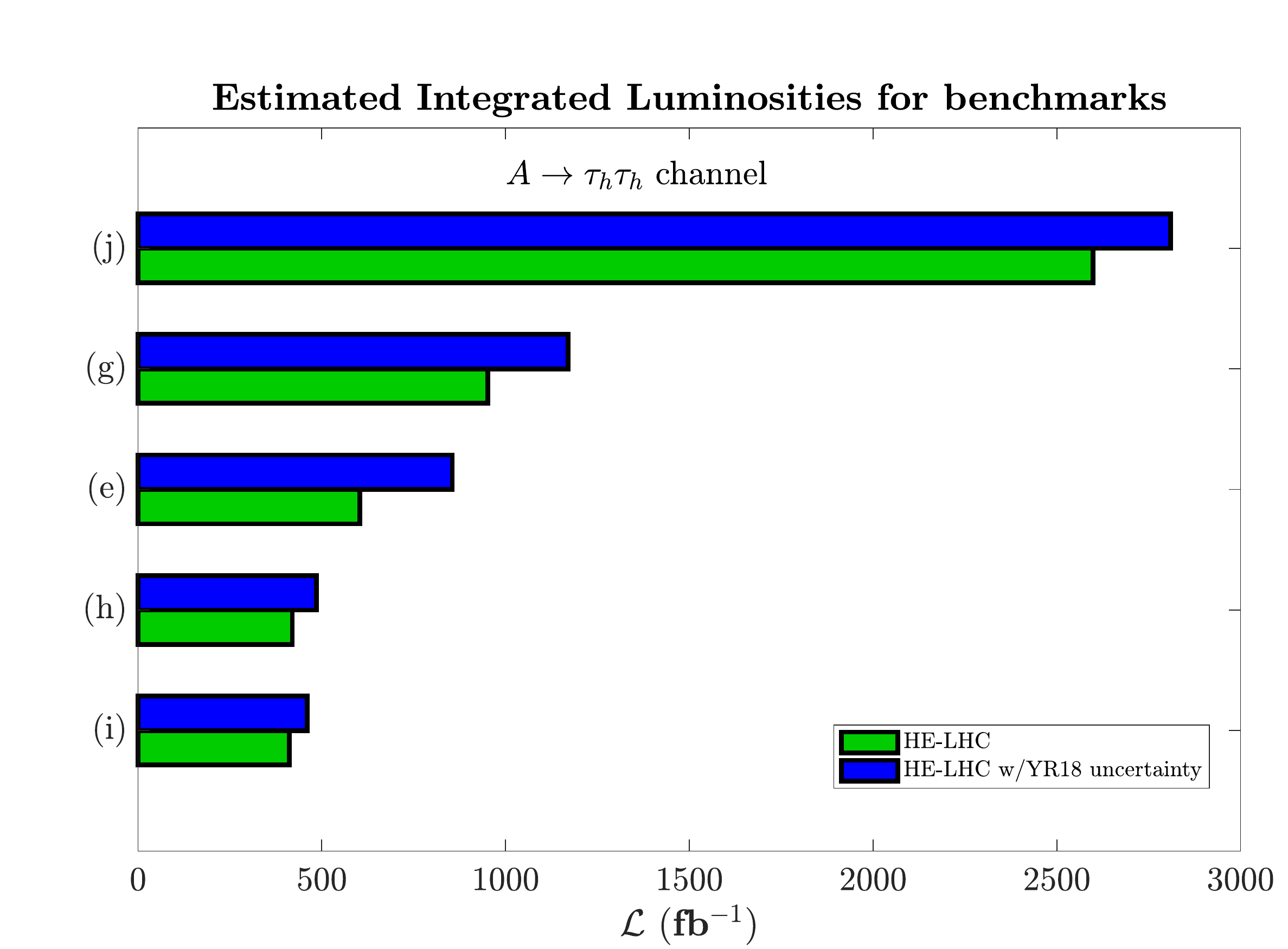}
      \caption{Left panel:  five benchmarks of Table~\ref{tab1} that are discoverable at both HL-LHC and HE-LHC 
      with and without       
       the ``YR18 uncertainties". Right panel: the  remaining five benchmarks of Table~\ref{tab3}       
        that are discoverable only at HE-LHC  with  and without 
             the ``YR18 uncertainties"~\cite{Aboubrahim:2019vjl}.}
	\label{fig6}
\end{figure}

\section{Conclusions}\label{sec:conc}

Dark matter does not need to reside in the visible sector and may well be in a hidden sector and possess very weak interactions with the MSSM fields. Such a possibility can arise in SUSY models with $U(1)_X$ extension leading to a DM candidate with very small scattering cross-section allowing it to escape direct detection experiments. However, the effect of such a candidate can be seen through the decay of visible sector particles into the hidden sector DM. In the first part of this work we considered the lightest stau to be the NLSP of the extended MSSM/SUGRA model while the LSP is the neutralino of the hidden sector obtained after extending the neutralino mass matrix to a $6\times 6$ mass matrix so that the hidden sector neutralinos interact with the visible sector via the small gauge kinetic and mass mixings. Due to this very small mixing and additional phase space suppression between the LSP and NLSP, the stau will decay to the hidden sector neutralino after having traveled a considerable distance in the ID owing to its long lifetime. Thus the stau may leave a track in the ID making it visible to experiments at HL-LHC and HE-LHC. We have shown that half of the selected benchmarks can be observed at HL-LHC while all benchmarks corresponding to a wider mass range of the stau may be visible at HE-LHC.

One can add to the hidden sector matter fields charged under $U(1)_X$ while neutral under $U(1)_Y$. This is of interest especially for MSSM/SUGRA models where the LSP is a higgsino. This particle has very efficient annihilation in the early universe leading to a small relic abundance. This leads the way to the possibility of having at least one extra component of DM to saturate the relic density. The second component can arise from a hidden sector. Thus in the second part of this work, we consider a multicomponent DM model consisting of a Majorana neutralino from the visible sector and a Dirac fermion from the hidden sector as the second component. It is shown that the Dirac fermion provides the dominant contribution to the relic density. The higgsino LSP in the visible sector can be obtained when the $\mu$ parameter is small which means that the MSSM heavy Higgs bosons can have masses under 1 TeV and thus accessible to the LHC. We study a set of benchmarks for the extended model and show that the CP odd Higgs boson in models of this type is observable when investigating the $\tau_h\tau_h$ final state at the HL-LHC and HE-LHC. It is seen that a CP odd Higgs with a mass up to 450 GeV for $\tan\beta\leq 12$ may be discoverable at HL-LHC. The discovery reaches 750 GeV at the HE-LHC with an integrated luminosity of $\sim 2600$ fb$^{-1}$ which is just a fraction of the optimal luminosity of 15 ab$^{-1}$ that HE-LHC can deliver. With the optimal luminosity the mass reach of HE-LHC for the CP odd Higgs mass will certainly extend far above 750 GeV.

\vspace{2cm}

\textbf{Acknowledgments: }
The analysis presented here was done using the resources of the high-performance  Cluster353 at the Advanced Scientific Computing Initiative (ASCI) and the Discovery Cluster at Northeastern University.  This research was supported in part by the NSF Grant PHY-1620575 and and PHY-19060675,

\end{document}